\documentstyle[epsf,epsfig,e-e-ijmpa,twoside]{article}

\renewcommand{\thefootnote}{\fnsymbol{footnote}}  

\def\rar{\rightarrow}

\def\epem{$e^+e^- \,$}
\def\emem{$e^-e^- \,$}
\def\alr{$A_{LR} \,$}
\def\alrm{A_{LR}}
\def\z0{$Z^0 \,$}
\def\sintw{\sin^2(\theta_W^{eff}) \,}
\def\pe{$P_e \,$}
\def\pelum{$P_e^{lum} \,$}
\begin{document}

\normalsize\textlineskip
\pagestyle{plain}
\begin{flushright}
{\small
SLAC--PUB--8397\\
March, 2000\\}
\end{flushright}
\vspace{.8cm}

\title{POLARIMETRY AT A FUTURE LINEAR COLLIDER -- \\
HOW PRECISE?}
\author{M. WOODS%
\footnote{Work supported by the Department of Energy, 
Contract DE-AC03-76SF00515.}}
\address{Stanford Linear Accelerator Center \\
Stanford University, Stanford, CA 94309, USA}
\maketitle\abstracts{
At a future linear collider, a polarized electron beam will play an important
role in interpreting new physics signals.  Backgrounds to a new physics 
reaction can be reduced by choice of the electron polarization state.  The 
origin of a new physics reaction can be clarified by measuring its 
polarization-dependence.  This paper examines some options for polarimetry with
an emphasis on physics issues that motivate how precise the polarization 
determination needs to be.  In addition to Compton polarimetry, the possibility
of using Standard Model asymmetries, such as the asymmetry in forward 
W-pairs, is considered as a possible polarimeter.  Both \epem and \emem 
collider modes are considered. 
}
\setcounter{footnote}{0}
\renewcommand{\thefootnote}{\alph{footnote}}
\vspace*{1pt}\textlineskip	

\vfill

\begin{center}
Presented at \\
\vspace{3mm}
{\it \boldmath$e^-e^- 99$ \\
3rd International Workshop on Electron-Electron Interactions at TeV Energies \\
December 10-12, 1999 \\
UC Santa Cruz, Santa Cruz, CA, USA}  \\
\end{center}

\pagebreak
\section{Introduction}

This paper addresses the issue of how accurately one should measure 
the electron beam polarization, \pe, at a future linear collider.  
The collider performance parameters shown in Table~\ref{tab:machine} are
used.   I begin by
considering how accurately Standard Model (SM) asymmetries may be measured, 
and the requirements these measurements place on the polarimetry.  In addition
to Compton polarimetry,\cite{Woods1} the measurement
of SM physics asymmetries for polarimetry is considered.  This type of 
polarimetry has previously been proposed when both
colliding beams are polarized.\cite{Blondel,Gambino}  Here, I additionally 
consider
the possibility to use the asymmetry in forward W-pairs in \epem\ collisions,
when only the electron beam is polarized.  

The precision of the polarimetry can
affect the discovery potential for a new physics signal.  I examine the 
importance of precise polarimetry for accurately assessing W-pair backgrounds.

Beam-beam effects in the collision process have a significant 
impact on polarim-etry.  First, significant depolarization can result,
and one needs to determine how the measured \pe
 is related to the luminosity-weighted polarization, $P_e^{lum}$.  At high
luminosities with large beamsstrahlung disruption, it may no longer be possible
to place a Compton polarimeter in the extraction line from the Interaction
Region (IR).  This will limit the ability to accurately determine the 
amount of depolarization in the collision process. 

\begin{table}[h]
\tcaption{Collider performance parameters for this study.}
\vspace{2mm}
\begin{center}
\begin{tabular}{ccc}
\hline
Parameter	& \epem		& \emem	\\
\hline
$\sqrt{(s)}$	& 500 GeV	& 500 GeV \\
$\int L\,dt$	& 80 fb$^{-1}$	& 25 fb$^{-1}$ \\
$P_1$		& 90$\%$	& 90$\%$ \\
$P_2$		& 0$\%$		& 90$\%$ \\
\hline
\end{tabular}
\end{center}
\label{tab:machine}
\end{table}

\section{Standard Model Asymmetries in \boldmath\epem}

\begin{figure} [t]
\begin{center}
\epsfig{figure=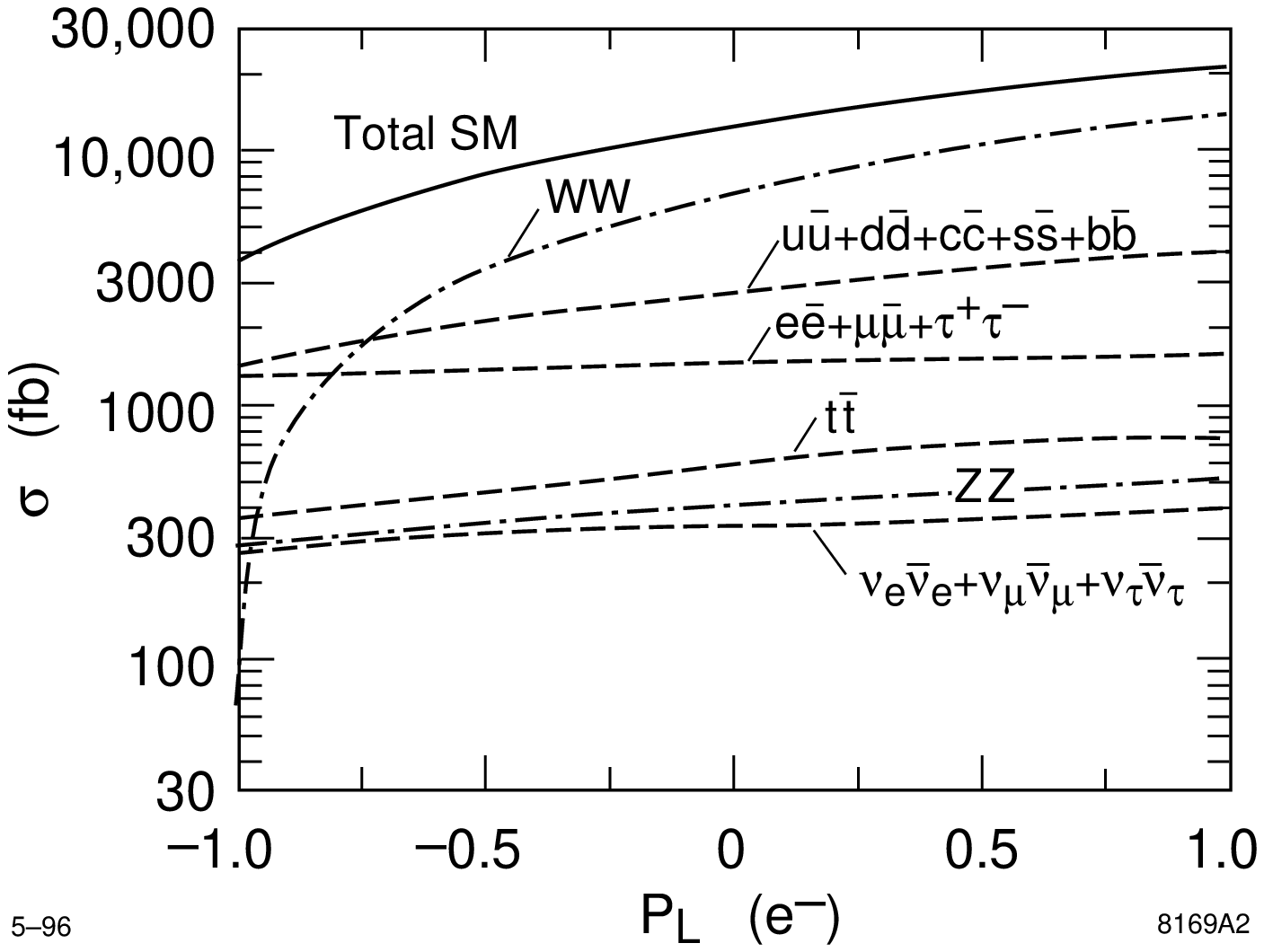,width=10cm}
\end{center}
\fcaption{Dependence of Standard Model cross sections on the electron beam
polarization at a 500 GeV \epem collider.\cite{snowmass96}}
\label{fig:smasym}
\end{figure}

For an \epem collider with 500 GeV center-of-mass energy, the dependence of
SM production cross sections on \pe is plotted in
Figure~\ref{fig:smasym}.  
This Figure is taken from a Report for Snowmass 
1996 on the {\it Next Linear Collider}\cite{snowmass96} and assumes the 
positron beam is unpolarized.  Following the study in that Report,  
some SM asymmetries are estimated for a detector with an acceptance of
$|cos\theta | < 0.99$, for an integrated luminosity of $80\,fb^{-1}$.
These are summarized in Table~\ref{tab:asymepem}.  The left-right asymmetry,
\alr, and its statistical and systematic uncertainties are given by
\pagebreak
\begin{eqnarray}
A_{LR} & = & {(\sigma_L-\sigma_R) \over (\sigma_L+\sigma_R)} \nonumber \\
\delta A_{LR} & = & \sqrt{{[1-(PA_{LR})^2] \over P^2N} +
		    \left( {\delta P \over P} \right)^2A_{LR}^2} \nonumber \\
	&	= & \sqrt{(\delta A_{LR}^{stat})^2 + (\delta A_{LR}^{syst})^2},
\nonumber
\end{eqnarray}
where $P=90\%$ is the electron polarization;\cite{Subashiev}
$N_L$ ($N_R$) is the number of
observed events with the left-(right-)polarized beam, and $N=N_L+N_R$.  Equal
integrated luminosities are accumulated with the left- and right-polarized
electron beam.

\begin{table} [h] 
\tcaption{Standard Model Production Asymmetries.}
\vspace{2mm}
\begin{center}
\begin{tabular}{cccc}
\hline
Final State	& \# events  & \alr  & ${\delta \alrm^{stat} \over \alrm}$ \\
\hline
$W^+W^-$	& 560K 	& 99.2$\%$  &  $0.0007$  \\
$q \overline{q}$ & 250K &  45$\%$  &  $0.005$ \\
$l^+ l^-$       & 120K  & 10$\%$   &  $0.032$ \\
\hline
\end{tabular}
\end{center}
\label{tab:asymepem}
\end{table}

\begin{figure} [t]
\begin{center}
\epsfig{figure=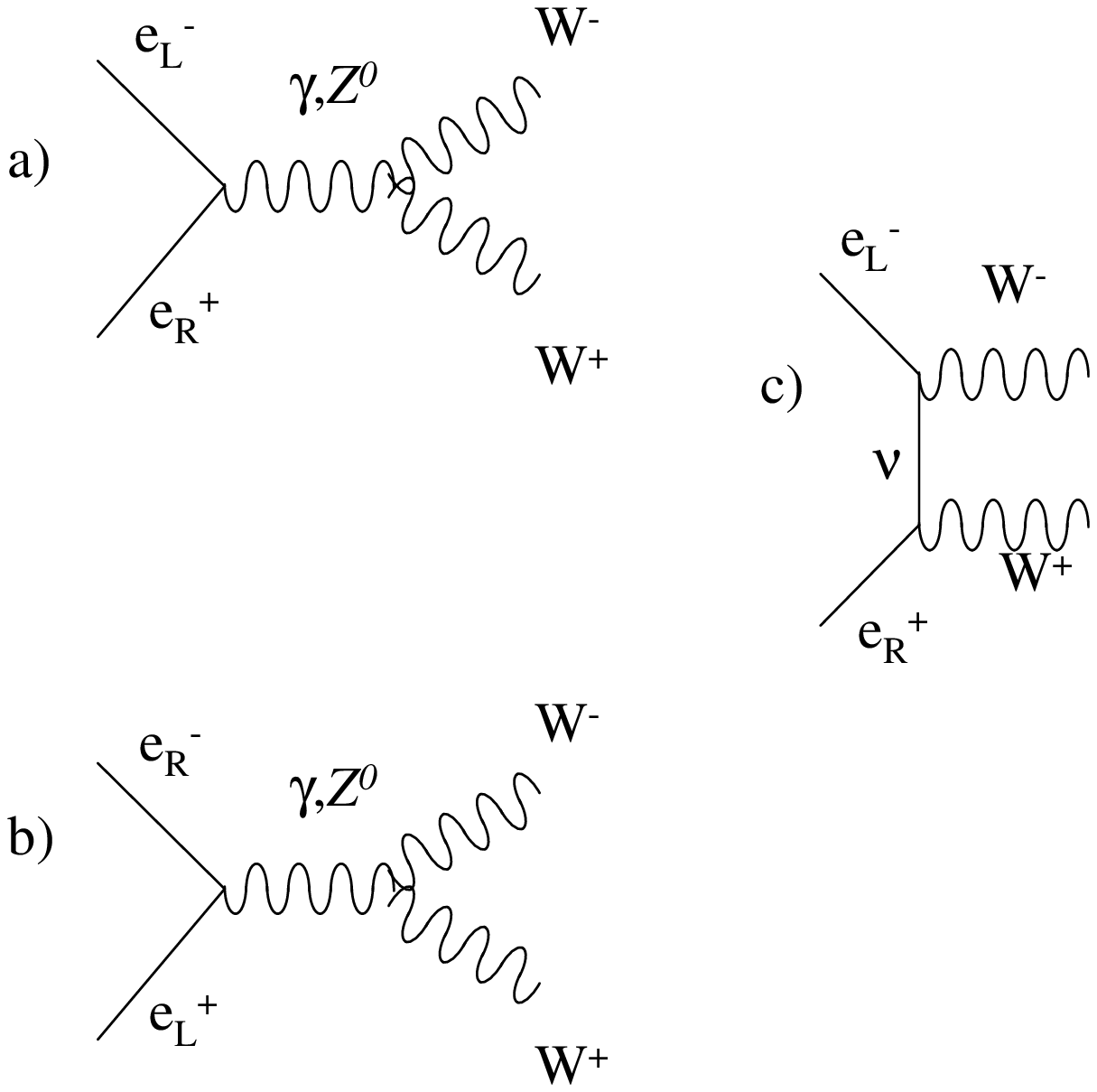,width=8cm}
\end{center}
\fcaption{Feynman diagrams for W-pair production at an \epem collider.}
\label{fig:feynman}
\end{figure}

Table~\ref{tab:asymepem} indicates that it is necessary to have better
than $1\%$ polarimetry to fully exploit testing SM predictions for these
asymmetry measurements.  This can be achieved with a precise Compton 
polarimeter.  It is also interesting to consider whether the 
asymmetry in W-pairs could be used as a polarimeter.  
The Feynman diagrams contributing to W-pair production are shown in 
Figure~\ref{fig:feynman}.  The diagram in Figure~\ref{fig:feynman}b is highly
suppressed.  In the forward detector regions ($\cos\theta > 0.7$), the 
exchange diagram in 
Figure~\ref{fig:feynman}c dominates and \alr for W-pairs is essentially
$100\%$.  The measured left-right asymmetry in forward W-pairs can therefore 
be used to determine the electron
beam polarization.\footnote{One will still be able to check for new physics
processes that might affect this polarization 
determination, by measuring the dependence of 
the result on the polar scattering angle.}    
\hspace{0.1cm}To achieve sub-1$\%$ accuracy for \pe will require achieving
backgrounds to the W-pair sample below $1\%$.  
This has been achieved at LEP200 for the W mass measurements,
where one of the Ws is required to decay to $\mu\nu_{\mu}$ and tight cuts
are placed on the reconstructed W mass.\cite{charlton}  
A detailed study needs to be done to determine whether this low background
could also be achieved with
the forward detector regions at a linear collider.  An advantage of using a 
detector physics asymmetry for polarimetry rather than Compton polarimetry is
that \pelum is directly determined and beam-beam
depolarization effects are properly accounted for.\footnote{If depolarization 
effects are significant, however, there may be a significant 
dependence of \pe on the electron's effective collision 
energy.  In this case, the energy distribution of collision electrons in the 
W-pair sample should be the same as the distribution of collision electrons in 
the physics sample being studied to ensure that \pelum is being accurately 
determined.}

A wonderful example of the physics possibilities with precise asymmetry
measurements is the linear collider \z0-Factory option.  It is very desirable
to accumulate a large \z0 sample ($>10$ million) with a polarized electron beam.
For example, Ref.~\citenum{monig} considers achieving a sample of $10^9$ \z0 decays
with $80\%$ electron polarization and $60\%$ positron polarization.  This 
enables the determination of the weak mixing angle, by measuring the
polarization-dependent cross sections for \z0 production,
with an unprecedented 
accuracy of $\delta \sin^2(\theta_W^{eff})=1.3 \cdot 10^{-5}$.  In this case,
the availability of a polarized positron beam allows for very precise 
polarimetry.\cite{Blondel}  For the case where the positron beam is not 
polarized, it will still be very desirable to make a significant improvement
on the SLD weak mixing angle measurement, 
$\delta\sintw=  2.8 \cdot 10^{-4}$.\cite{sld}  This will
require better than $0.5\%$ polarimetry from a Compton polarimeter located 
in the extraction line from the IR.  One can also hope to resolve whether the
anomalies observed in the $Zb\overline{b}$ asymmetry measurements\cite{sld}
at SLD and LEP are due to statistical 
fluctuations, systematic problems or new physics.  

\section{Standard Model Asymmetries in \boldmath\emem}

For the linear collider operating with \emem collisions, both beams can be 
polarized.  One of the important measurements that will be made is an accurate
measurement of the weak mixing angle away from the \z0-pole, by measuring 
the polarization-dependent cross sections in Moller scattering
($e^-e^- \rar e^-e^-$).  With both beams
polarized, one can measure three independent asymmetries which can be chosen 
to be
\begin{eqnarray}
A_1 & = & {N_{LL}-N_{RR} \over N_{LL}+N_{RR}} \nonumber \\
A_2 & = & {N_{RR}-N_{LR} \over N_{RR}+N_{LR}} \nonumber \\
A_3 & = & {N_{LR}-N_{RL} \over N_{LR}+N_{RL}}. \nonumber 
\end{eqnarray}
From these asymmetry measurements, one can determine $\sintw, P_1$, and $P_2$.
A detailed study of this has been done in Ref.~\citenum{Gambino} for a 500 GeV collider,
beam polarizations of 90$\%$, detector acceptance with $|\cos\theta|<0.995$,
and integrated luminosity of 25 $fb^{-1}$.  They find that the beam 
polarizations can be determined with an accuracy of 0.9$\%$, and that the
weak mixing angle can be determined with an accuracy of 
$\delta \sintw=0.00026$.  This accuracy is comparable to that achieved with
SLD's \alr measurement at the \z0-pole and will be the best measurement of 
the weak mixing angle away from the \z0-pole.  The running of 
$\sintw$\cite{Marciano}  with
$Q^2$ will be measured with excellent precision, a factor 3 better than 
that expected from the SLAC E158 experiment.\cite{e158}  Excellent sensitivity 
to additional \z0 bosons (up to $m_{Z'} \approx 10$ TeV) and to electron
compositeness (up to a compositeness scale, $\Lambda \approx 100$ TeV) will be 
achieved.  The beam polarization
uncertainty is comparable to what one can expect with a Compton polarimeter,
and has the advantage that it directly measures $P_e^{lum}$.  The determination
of \pelum from the Moller scattering analysis can also be applied to 
other physics
analyses.\footnote{One may have to make small corrections for the dependence
of \pelum on the
energy distribution of the collision electrons if depolarization effects are
significant.}    

\pagebreak
\section{Background Suppression of W-pairs}

For the \epem collider, W-pair background will be an obstacle for observing
new physics reactions.  Beam polarization will be an important tool for 
understanding and reducing this background.  The production cross section
for W-pairs may be written as
\begin{eqnarray}
\sigma(P_1,P_2)={1 \over 4}[(1-P_1)(1+P_2)\sigma_{LR}+
	(1+P_1)(1-P_2)\sigma_{RL}], \nonumber 
\end{eqnarray}
where $P_1$ is the electron beam polarization; $P_2$ is the positron beam
polarization; and $\sigma_{LR}$ ($\sigma_{RL}$) is the W-pair production
cross section 
for a left-(right-)handed electron colliding with a right-(left-)handed
positron.  As noted in Ref.~\citenum{Omori}, 
$\sigma_{RL}$ is highly suppressed, 
$(\sigma_{RL} / \sigma_{LR} \approx 0.004)$.  

For assessing the utility of achieving high beam polarizations, it is useful
to construct a Figure-of-Merit (FOM) defined as the ratio,
$R=\sigma(P_1,P_2)^{max} / \sigma(P_1,P_2)^{min}$ where the maximum 
(minimum) W-pair
production cross section is achieved with a left-(right-)polarized electron 
beam and a right-(left-)polarized positron beam.  Table~\ref{tab:fom}
summarizes this FOM for some possibilities for the beam polarizations.
It is desirable to achieve a high
electron beam polarization, since the FOM increases by a factor of 2 when 
improving $P_1$ from $80\%$ to $90\%$.  The utility of polarizing the positron 
beam is also evident,\cite{Omori} though this may be difficult to
implement in a cost-effective way.

The accuracy of the polarization determination will be important
for assessing the suppression of the W-pair backgrounds.  To illustrate this,
consider a potential experiment where the electron beam polarization is 
$90\%$ and the 
positron beam is unpolarized.  Suppose an analysis for isolating a new
physics signal yields 400 candidate events after analysis cuts, but with
no cut on the polarization state.  If the right-polarized electron
state is chosen, suppose 40 events are observed to survive.  This would be an 
excess of 20 events over what would be expected if the entire sample were due
to W-pair backgrounds.  The measured left-right asymmetry would be
$$ A_{LR}^{meas} = 0.80 \pm 0.024(stat) \pm {\delta P \over P}. $$
The uncertainty on $A_{LR}^{meas}$ is summarized in Table~\ref{tab:expt}
for 3 possible values of the accuracy of the polarization determination.
To achieve a $4\sigma$ signal will require better than $1\%$ polarimetry.
More precise polarimetry is generally desirable as the beam polarization
increases, to assure accurate assessment of the W-pair background. 

\begin{table} [t]
\tcaption{Figures-of-Merit for W-pair Background Suppression.}
\vspace{2mm}
\begin{center}
\begin{tabular}{ccc}
\hline
Electron Polarization	& Positron Polarization  & FOM \\
\hline
0	&	0	&	1.0	\\
0.8	&	0	&	8.7	\\
0.9	&	0	&	17.7	\\
0.8	&	0.5	&	24.5	\\
0.8	&	0.8	&	61.8	\\
0.9	&	0.8	&	103.2	\\
1.0	&	0	&	260.	\\
1.0	&	1.0	&	260.	\\	
\hline
\end{tabular}
\end{center}
\label{tab:fom}
\end{table}
\begin{table} [t]
\tcaption{Dependence of Signal Asymmetry Error on Polarization Error.}
\vspace{2mm}
\begin{center}
\begin{tabular}{ccc}
\hline
${\delta P \over P}$	& $\delta A_{LR}^{meas}$ & Significance \\
			&			 & of Result \\
\hline
0	&	$2.4\%$ & $4.2\sigma$ \\
$1\%$	&	$2.6\%$ & $3.8\sigma$ \\
$2\%$	&	$3.1\%$ & $3.2\sigma$ \\
\hline
\end{tabular}
\end{center}
\label{tab:expt}
\end{table}

\section{Beam-beam Effects on Precise Polarimetry}

Beam-beam effects in the collision process can cause significant depolarization
due to spin precession and the Sokolov-Ternov spin flip mechanism.\cite{Chen}
For a 500 GeV NLC with a luminosity of $6\cdot10^{33}$ cm$^{-2}$s$^{-1}$, 
the luminosity-weighted depolarization is estimated to be about $0.15\%$.
For a 1 TeV NLC with a luminosity of $1.4\cdot 10^{34}$ cm$^{-2}$s$^{-1}$,
it is estimated to be about $1.5\%$.\cite{nlcmemo}  These calculations
should be checked experimentally, however.  This can be done with a Compton
polarimeter in the extraction line from the IR, by comparing polarization
measurements with and without collisions.\cite{Woods1}  

The extraction line Compton polarimeter measures the total depolarization
in the collision process.  The luminosity-weighted depolarization is
typically one-quarter of this.\cite{Chen}  This is easily understood for 
spin precession effects, where the
depolarization has a quadratic dependence on the precession angle and one 
assumes
that half the precession occurs before the hard 
collision.\footnote{This 
assumption may not be valid if the beams undergo significant betatron 
oscillations during the collision.  In that case, the luminosity-weighted
depolarization may be comparable to the total depolarization.}  
\hspace{0.1cm}For example, 
the depolarization due to 
the large disruption angles of the beams is
\begin{eqnarray}
\Delta P_e & = & {1 \over 2} \left(\gamma {g-2 \over 2}\right)^2 
\left[\sigma_{x'}^2+\sigma_{y'}^2\right] \nonumber \\
\Delta P_e^{lum} & \approx & {1 \over 2} \left(\gamma {g-2 \over 2}\right)^2 
\left[\left({\sigma_{x'} \over 2}\right)^2 + \left({\sigma_{y'} 
\over 2}\right)^2\right] \nonumber \\ 
\Delta P_e^{lum} & \approx & {1 \over 4} \Delta P_e, \nonumber  
\end{eqnarray}
where $\sigma_{x'}$ ($\sigma_{y'}$) is the disrupted x (y) angular divergence;
$\gamma({g-2 \over 2})$ is the spin precession factor; $\Delta P_e$ is the 
total depolarization and $\Delta P_e^{lum}$ is the luminosity-weighted
depolarization.  

It is important for IR physicists to include 
depolarization in their tabulations of beam-beam effects.
Extensive tables of beam-beam effects for energy distributions, luminosity
distributions and outgoing angular distributions are produced in the design
reports for NLC, JLC and TESLA.  Surprisingly, depolarization effects are not
included.  This is presumably due to the lack of depolarization calculations
in the beam-beam effect simulation codes used,
CAIN\cite{cain} and GUINEA-PIG.\cite{guinea}  These
programs should be improved to include depolarization effects, and 
depolarization should 
be included as a key element in tables summarizing beam-beam effects. 

At high luminosities, or for the more severe beam disruption experienced in
\emem collisions, it becomes difficult to transport the disrupted beam 
cleanly to the beam dumps.\cite{yuri}  This may lead to an extraction line 
design that
precludes a Compton polarimeter and other beam diagnostics.  In this 
scenario, one will have to rely on a Compton polarimeter before the IP, and 
possibly utilize SM physics asymmetries for polarimetry as well.  By
comparing the upstream Compton polarization measurement with that provided by
a SM physics asymmetry, the beam-beam depolarization can be determined.  This
depolarization determination will have greater systematic uncertainty than that
achievable with an extraction line Compton polarimeter, for which many sources 
of systematic error will cancel when determining the amount of depolarization.
However, \pelum and $\Delta P_e^{lum}$ will be more directly determined, 
perhaps offsetting this disadvantage.  Ideally, both extraction line Compton 
polarimetry and SM physics asymmetry polarimetry will be achievable.

\section{Conclusions}

Precise measurements of SM asymmetries in both \epem and 
\emem collider modes require better than $1\%$ polarimetry.  Sub-$1\%$
polarimetry may also be required to accurately assess W-pair backgrounds in a 
discovery search for a new physics signal at an \epem collider.

A Compton polarimeter in the extraction line from the IP is desirable,
especially for its ability to accurately measure depolarization effects.
SM physics asymmetries are useful for polarimetry when both colliding 
beams are polarized.  For the \epem collider with only the electron beam 
polarized, the asymmetry in forward W-pairs may also prove useful as a 
polarimeter.  For the \z0-factory \epem collider (below W-pair threshold) 
with no positron polarization, a very precise Compton polarimeter in the 
extraction line is required. 

There is a need to include depolarization in the tables summarizing 
beam-beam effects.  Including depolarization calculations in 
the simulation programs for beam-beam effects will assist this. 

\nonumsection{Acknowledgements}

I would like to thank Michael Peskin for discussions and comments regarding
this paper.  I would also like to thank Clemens Heusch for organizing this
meeting and for promoting the less conventional aspects of linear colliders.

\pagebreak
\nonumsection{References}

\end{document}